\documentclass[12pt]{article} \usepackage{graphicx}
\usepackage{amssymb}

\newcounter{mnotecount}[section] \newcommand{\mnote}[1]
{\protect{\stepcounter{mnotecount}} $^{\mbox{\footnotesize
mn:\themnotecount}}$ \marginpar{\raggedright\tiny\em
\hspace*{-2em}$\bullet$\themnotecount: #1}}

\newtheorem{Theorem}{Theorem}
\newtheorem{Proposition}[Theorem]{Proposition}
\newtheorem{Definition}[Theorem]{Definition}
\newtheorem{Lemma}[Theorem]{Lemma}
\newtheorem{Corollary}[Theorem]{Corollary}
\newcommand{\Proof}{\par\medskip\noindent {\em Proof:}\ \ }
\newcommand{\Remark}{\par\noindent{\sc{Remark:}\ \ }}

\newcommand{\QED} {\hfill$\hbox{\vrule height1.3ex width1.3ex
depth.1ex}\ $
    \par\smallskip}
\newcommand{\bref}[1]{(\ref{#1})}

\newcommand{\cut}%
{\hbox{%
\,{\vrule height.1ex width.7ex depth.1ex}%
{\vrule height1.3ex width.1ex depth.1ex}\,}}

\DeclareFontFamily{OT1}{rsfs}{} \DeclareFontShape{OT1}{rsfs}{m}{n}{
<-7> rsfs5 <7-10> rsfs7 <10-> rsfs10}{}
\DeclareMathAlphabet{\mycal}{OT1}{rsfs}{m}{n}
\newcommand{\bR}{{\mathbb{R}}}

\newcommand{\tfrac}[2]{{\textstyle{\frac{#1}{#2}}}}
\newcommand{\half}{\tfrac{1}{2}} 

\newcommand{\ddR}{\frac{d}{dR}} 

\newcommand{\trg}{\textrm{tr}_\gamma}
\newcommand{\newton}{\textrm{\sc{g}}\;}

\title{Spherically Symmetric Dynamical Horizons} 

\author{Robert Bartnik\thanks{%
School of Mathematical Sciences, Monash University, Victoria, 3800
Australia.  E-mail: robert.bartnik@sci.monash.edu.au}
\ and James Isenberg\thanks{%
Department of Mathematics and Institute for Theoretical Science,
University of Oregon, Eugene, OR 97403
USA. E-mail:jim@newton.uoregon.edu} }

\begin{document}
\maketitle

\begin{abstract}
  We study spherically symmetric dynamical horizons (SSDH) in
  spherically symmetric Einstein/matter spacetimes.  We first
  determine sufficient and necessary conditions for an initial data
  set for the gravitational and matter fields to satisfy the dynamical
  horizon condition in the spacetime development. The constraint
  equations reduce to a single second order linear ``master''
  equation, which leads to a systematic construction of all SSDH
  initial data sets with certain boundedness conditions.  Turning from
  construction to existence, we find necessary and sufficient
  conditions for a given spherically symmetric spacetime to contain a
  SSDH.


\end{abstract}

\section{Introduction} 

Since their introduction four years ago by Ashtekar and his
collaborators \cite{AK02} \cite{AK04}, dynamical horizons have been
found to be very useful for the study of the dynamical formation of
black holes. Among other things, they have been used to study
gravitational wave fluxes and their influence on the areas of black
holes as they develop, and they have been used in building analogues of
black hole thermodynamics for astrophysical systems evolving into
black holes \cite{AK04}.

How extensively do dynamical horizons occur in solutions of Einstein's
equations? While there have been studies, both numerical and
analytical, of certain families of explicit examples \cite{BBGV05}, no
attempt has yet been made to determine the parameter space of
solutions which admit them. We do this here for the special case of
spherically symmetric solutions.

We obtain two types of results, on the \emph{construction} of
spacetimes with dynamical horizons (by constructing dynamical horizon
initial data), and on \emph{locating} dynamical horizons in given
spacetimes. 

We first consider the problem of
constructing initial data satisfying both the dynamical horizon
condition, and the Einstein constraint equations.  Proposition
\ref{prop1} parametrizes those initial data on $I \times S^2$ which
are spherically symmetric, which solve the Einstein constraint equations,
and which serve as a dynamical horizon for their spacetime development. We
find that the collection $\mathcal{D}$ of such initial data is
parametrized by two free functions ($\xi,\tau$) on an interval and a
pair of real constants ($y_0,y_1$).  That is, for every choice of
these two functions, we can generate an explicit initial data set of
this sort; conversely, every data set in $\mathcal{D}$ can be
generated in this way.  For simplicity we consider only bounded
potentials, but there are many examples of dynamical horizons with
unbounded potentials \cite{AK02,BBGV05}, whose global properties can
be studied using similar techniques.

Note that since spherically symmetric solutions of the vacuum Einstein
equations contain no dynamical horizons\footnote{By Birkhoff's theorem
  the vacuum region is locally Schwarzschild, which has vanishing
  outer expansion only on the horizons, which are not
  spacelike.}, we consider here the Einstein
constraint equations with generic matter, represented by an energy
density field $\rho$ and a radial momentum density field $\xi$.

Second, we consider the problem of finding dynamical horizon(s) in a
given (smooth, globally hyperbolic) spacetime.  Lemma \ref{lem1} and
Proposition \ref{prop3} give conditions on the stress-energy tensor
which lead naturally to a general existence result for
dynamical horizons in the spacetime.  

Our arguments depend crucially on two quite interesting and novel
identities: a constraint \bref{7} on two components of the
stress-energy tensor, and a reformulation of the spherically symmetric
constraint equations as a linear second order ordinary differential
equation \bref{yeq} for the volume function $y=r^3$.  We expect these
expressions will play an important role in extending our results to
non-spherical spacetimes.

\section{Preliminaries}

We work here with four dimensional spacetimes $(M, g, T)$ satisfying
the Einstein gravitational field equations $G_{\mu \nu}= 8 \pi \newton
T_{\mu \nu}$, where $M$ is a spacetime manifold, $g$ is a Lorentz
signature metric invariant under an $SO(3)$ action and with Einstein
curvature tensor $G_{\mu \nu}=R_{\mu\nu}-\half R g_{\mu\nu}$,
$T_{\mu\nu}$ is a stress-energy tensor field which may or may not be a
functional of a specific set of field variables, and $\newton$ is the
Newton gravitational constant. In our spacetime results (Section
\ref{sec4}) we require that $(M, g, T)$ satisfy an energy condition;
in our initial data results (Section \ref{sec3}), such a condition is
not needed.

\begin{Definition}
An embedded hypersurface $\Sigma^3$ in a spacetime $(M,g,T)$, is a
\emph{dynamical horizon} for the spacetime if the following conditions
are met:
\begin{itemize}
\item[(DH1)] $\Sigma^3$ is spacelike;
\item[(DH2)] $\Sigma^3$ is foliated by \emph{marginally outer-trapped}
  2-surfaces, i.e., surfaces such that the expansion of the outer future null vector field
  $\ell=\ell_+$  vanishes,   $\theta_\ell=\theta_+ =0$;
\item[(DH3)] the marginally outer trapped 2-surfaces are \emph{inner
    trapped}, $\theta_n=\theta_-<0$, where $n=\ell_-$ is the inward
    future null vector field normal\footnote{normalised by
  $g(n,\ell)=-2$.} to the 2-surfaces.
\end{itemize}
\end{Definition}
A 2-surface with null normal expansions satisfying (DH2) and (DH3) is
said to be \emph{marginally trapped}.  A hypersurface satisfying just
(DH2) and (DH3) is called a \emph{marginally trapped tube}
\cite{AG05}; however here we are concerned only with the dynamical
horizon case.  Under spherical symmetry we may assume the 2-surfaces
are isometric 2-spheres of positive radius.

A spacetime may generally contain a number of dynamical horizons.
However, as shown in \cite{AG05}, the foliation of a given dynamical
horizon by marginally trapped surfaces is unique. It also follows from
the results in \cite{AG05} that if a given spacetime satisfies the
null energy condition (i.e., $T(\ell,\ell)\geq 0$ for every null
vector $\ell$), then no spacetime region admits a local foliation by
dynamical horizons. 

\section{Constructing Initial Data} \label{sec3}
In this section, we derive an algorithm for the construction of
spherically symmetric initial data sets which satisfy both the
constraint equations and the dynamical horizon condition. In doing so
we determine the freely specifiable parts of such initial data,
showing that there is a bijective correspondence between this free
data on the one hand, and spherically symmetric dynamical horizon
solutions of the constraints on the other hand. We then describe some
example solutions, and we discuss some physically motivated
restrictions one might impose on these solutions.


\subsection{Construction Algorithm and the Free Data}
An initial data set for the  Einstein equations consists of
$(\Sigma^3, \gamma, K, \rho, J)$, where $\Sigma^3$ is a three
dimensional manifold (the initial slice), $\gamma$ is a Riemannian
metric (the initial metric), $K$ is a symmetric tensor (the initial
extrinsic curvature), $\rho$ is a non-negative function ($8 \pi
\newton$ times the initial matter energy density), and $J$ is a
one form field ($8 \pi \newton$ times the initial momentum
density).  Assuming this data satisfies the Einstein
constraint equations,
\begin{eqnarray}
\label{HamConstr}
2\rho &=& R_\gamma-K_{ab}K^{ab} + (\trg K)^2,
\\
\label{MomConstr}
J_a &=& \nabla^b K_{ab} - \nabla_a (\trg K),
\end{eqnarray}
where $R_\gamma$ is the scalar curvature of $\gamma$, it follows
(assuming appropriate matter evolution equations) \cite{FB52} \cite{W84}
\cite{CBG69} that there is a unique spacetime $(\Sigma^3 \times I, g,
T)$ which i) satisfies the Einstein gravitational and matter field
equations, ii) induces the initial data $(\Sigma^3, \gamma, K, \rho,
J)$ on the embedded hypersurface\footnote{in the sense that $\gamma$ is
  the pullback of $g$ to $\Sigma^3 \times \{0\}$, $K$ is the second
  fundamental form corresponding to $\Sigma^3 \times \{0\}$, $\rho=8
  \pi \newton T(e_\perp, e_\perp) $ for $e_\perp$ the future-pointing
  timelike normal to $\Sigma^3 \times \{0\}$, and $J =8 \pi \newton
  T(e_\perp,\cdot )$} $\Sigma^3 \times \{0\}$, iii) is globally
hyperbolic, and iv) contains (up to diffeomorphism) all other
spacetimes which satisfy conditions i)-iii). This spacetime is called
the \emph{maximal spacetime development} of $(\Sigma^3, \gamma, K, \rho, J)$.
These results allow us to focus on the simpler initial data constraint
system, rather than confront the challenges of the full evolution
equations. Note, however, that if we are given a pair of solutions of
the constraint equations, it is generally not easy to determine
whether or not their spacetime developments are diffeomorphic.

If we restrict our attention to initial data sets which are
spherically symmetric, with $\Sigma^3$ diffeomorphic to the three
dimensional annulus $I \times S^2$, then we may write
\begin{eqnarray}  \label{gamma}
\gamma &=& dR^2 +r^2(R)( d\theta^2+ \sin^2\theta\, d\phi^2) ,
\\
 \label{K}
K  &=& \tfrac{1}{2} \mu(R) \gamma +(\tau (R)-\tfrac{3}{2}\mu(R))\,dR^2,
\end{eqnarray}
and $J=\xi(R)\, dR$, where $R \in (R_0, R_1)$, $\theta \in (0,\pi)$ and
 $\phi\in(0, 2\pi)$). (Note that the coordinate $R$ measures radial geodesic distance, and $r(R)$ is
 the spherical radius function.) Hence the full set of (spherically symmetric)
initial data is parametrized by five real valued functions: $r(R),
\mu(R), \tau(R), \rho(R)$ and $\xi(R)$. Note that $\tau=\trg K$ is the
mean curvature of the spacetime hypersurface $\Sigma$, whilst $\mu$ is
the mean curvature in $\Sigma$ of the 2-spheres $S^2_r$ of constant
radius $r$.

The constraint equations
(\ref{HamConstr},\ref{MomConstr}) for spherical data take the form
\begin{eqnarray}
\label{ham}
2r  \frac{d^2r}{dR^2}& = & {}-\left(\frac{dr}{dR}\right)^2  +1 -
r^2(\tfrac{3}{4} \mu^2- \mu \tau + \rho)\,,
\\
\label{mom}
r \frac{d\mu}{dR} &=& (2\tau-3\mu) \frac{dr}{dR} -  r \xi \,.
\end{eqnarray}
We wish to study spherically symmetric solutions of the constraint
equations which serve as dynamical horizons for their spacetime
developments.  The dynamical horizon condition $\theta_+=0$ takes the form
\begin{equation}
\label{DH}
\theta_\ell = \theta_+ = \mu+\frac{2}{r} \frac{dr}{dR} =0,
\end{equation}
where $\mu$ is the trace of the extrinsic curvature $K$ over the $S^2$
tangent 2-planes.  Substituting for $\mu$ in \bref{ham} and \bref{mom}
gives two equations for $\tfrac{d^2r}{dR^2}$, and eliminating this
common term gives
\begin{equation}
\label{7}
\rho+\xi=\frac{1}{r^2}.
\end{equation} 
We thus obtain a somewhat peculiar, quite explicit, condition on the
matter field part of the initial data which must be satisfied if the
data set is to correspond to a dynamical horizon.

Before stating the main result on the existence and parameterization
of initial data, we note another curious property: substituting
\bref{7} and \bref{DH} in \bref{ham} leads to a linear ``master''
equation for $y:=r(R)^3$:
\begin{equation}
\label{yeq}
\frac{d^2y}{dR^2} + \tau \frac{dy}{dR} - \tfrac{3}{2}\xi y = 0\,.
\end{equation}
\begin{Proposition}\label{prop1}
  Suppose we are given functions $\tau,\xi\in C^0(\bR)$ and a
  solution $y=y(R)$ of \bref{yeq} on an interval $I=(R_1,R_2)$ on which
  $y$ is everywhere positive.  Defining $r = y^{1/3}$, $\mu$ by
  \bref{DH} and $\rho$ by \bref{7} gives a solution of the spherical constraint
  equations (\ref{gamma}--\ref{mom}).  Conversely, (\ref{7},\ref{yeq})
  hold if the given data $(r,\mu,\tau,\rho,\xi)$ satisfy
  (\ref{gamma}--\ref{mom}) and \bref{DH} on an interval where $r$ is
  everywhere positive.
  
  Furthermore, let $y=y(R)$ be the solution of \bref{yeq} with initial
  conditions
   \begin{equation}
   \label{yic}
     y(R_0) = y_0 >0 , \qquad \ddR y(R_0) = y_1 >0 ,
   \end{equation}
   and denote by $I=(R_1,R_2)$, $R_0\in I$, the maximal interval on
   which the condition $r>0$ holds.  Let $(y,r,\mu,\tau,\rho,\xi)$ be
   the corresponding solution of (\ref{gamma}--\ref{yeq}).  If
   $R_1>-\infty$ then we may normalise $R$ so $R_1=0$ and then
   $\lim_{R\downarrow 0}r(R)=0$ and $\gamma$ is singular at $R_1$;
   likewise if $R_2<\infty$ then $\lim_{R\uparrow R_2}r(R)=0$ and
   $\gamma$ is singular at $R_2$.

  The solution $(y,r,\mu,\tau,\rho,\xi)$ on the interval $(R_1,R_2)$
  specifies a dynamical horizon if additionally $\ddR y>0$ on $(R_1,R_2)$.
\end{Proposition}
\Proof Direct computation verifies the equivalence of the spherically
symmetric constraint equations (\ref{gamma}--\ref{mom}) to the linear
equation \bref{yeq}, augmented by formulas (\ref{DH}) for $\mu$, and
(\ref{7}) for $\rho$. This proves the first part of the proposition.

The continuity assumptions $\tau,\xi\in C^0(\bR)$ ensure that the
initial value problem for the linear ODE \bref{yeq} together with the
the initial conditions \bref{yic} has a unique global solution $y\in
C^2(\bR)$ which is positive in a neighborhood of $R_0$; hence
$\lim_{R\downarrow R_1} y(R)$ exists and the solution $y(R)$ extends
to $R<R_1$.  By maximality of $I$ it follows that either $R_1=-\infty$
or $R_1>-\infty$ and $y(R_1)=0$.  Assuming now that $R_1>-\infty$, we
may without loss of generality set $R_1=0$.  Uniqueness of solutions
of linear ODEs shows that $y_1 = \ddR y(0) > 0$, since we require
that $r$ is nowhere vanishing on $I$. Expanding in a Taylor series
about $R=0$ gives $y(R) = r^3(R) = y_1 R +O(R^2)$, where the error
term is controlled by $||\tau||_\infty +||\xi||_\infty$.  Thus $\gamma
= dR^2 +(R^{2/3}y_1^{2/3}+O(R^{4/3}))\, (d\theta^2+\sin^2\theta
\,d\phi^2)$ $= ((3/y_1)^2r^4+O(r^7))dr^2+ r^2 (d\theta^2+\sin^2\theta
\,d\phi^2)$, so the metric is singular at the symmetry centre $r=0$.
Thus although the radial equation \bref{yeq} is satisfied globally,
the physical solution does not extend to points where $r=0$. This
holds for $R_2$ as well as $R_1$. 

Finally, since 
\begin{equation}
\theta_- = \mu - \frac{2}{r}\ddR r = - \frac{4}{r} \ddR r =
- \frac{4}{3} \ddR \log y
\end{equation}
when $\theta_+=0$, the dynamical horizon condition (DH3) is
equivalent to $\ddR r>0$ and $\ddR y>0$.  Observe that (DH3) serves
only to restrict the solution to the sub-interval $(R_1,R_3)\subset
(R_1,R_2)$ on which $\ddR y>0$; without it the above solution could be
extended beyond $R_2$, to regions where $\theta_-\ge0$.  \QED
 




\Remark The dynamical horizon metric behaviour $\gamma\sim
c^2r^4dr^2+r^2 d\Omega^2$ is similar to that of a constant mean
curvature hypersurface near a point singularity \cite{Bartnik89b}.
However, \bref{7} shows the spacetime curvature has components
(ie.~$G(e_\perp,\ell)$) which are unbounded as $r\to 0$ and we show in
Corollary \ref{cor1} that $r=0$ is a true spacetime curvature singularity. 


\subsection{Examples}
As outlined above, the process of producing spherically symmetric
initial data sets which satisfy the constraints and the dynamical
horizon conditions from a given pair of freely chosen functions is
very straightforward. We now illustrate this process with some
examples.

\paragraph{Maximal Data with Co-Moving Matter:}
Setting $\xi=0$ corresponds to choosing the matter to be
co-moving with the surface orthogonal observers,  
while setting $\tau=0$ results in the mean curvature
of the initial data set vanishing, so the data is maximal.  With these
choices, equation \bref{yeq} becomes simply $ y''=0$, which has the
general solution $y(R)=y_1(R-R_0)+y_0$. In this case, we may without loss of generality choose $R_0=0, y_0=0$, and $y_1=\alpha^3$ (for $\alpha>0$) so $y(R)=\alpha^3 R$. It then follows that $r(R) =
\alpha R^{\frac{1}{3}}$, and consequently the metric takes the
  form
\begin{equation}
\gamma = dR^2 + \alpha^2 R^{\frac{2}{3}} d\Omega^2,
\end{equation}
where $d\Omega^2$ indicates the round sphere metric. Calculating $\mu$
from \bref{DH} and substituting into the formula for $K$, we have
\begin{equation}
K=\frac{2}{3\alpha^3R } dR^2 - \frac{1}{3 \alpha R^{\frac{1}{3}}}d\Omega^2.
\end{equation}
Calculating $\rho$ from \bref{7}, we have $\rho(R)=
\frac{1}{\alpha^2R^{\frac{2}{3}}}$.  We note that for this example,
the interval on which the solution is regular is $I=
(0, + \infty)$. The solution  is singular at $R=0$, where $r=0$. We also note that since $\frac{dy}{dR}=\alpha^3 >0$, the entire solution satisfies condition (DH3).

\paragraph{CMC Data with Co-Moving Matter:}
 We again set $\xi=0$, but set $\tau$ equal to a non
zero constant. The general solution\footnote{In this case, since the solution is not linear in $R$, setting $y_0=0$ is a restriction. So we let $y_0$ and $y_1$ be any pair of positive constants.} to the equation for $y$ is $y(R)
=\frac{y_1}{\tau}(1-e^{-R \tau})+y_0$,  which results in
\begin{equation}
\gamma = dR^2 +( \frac{y_1}{\tau}(1-e^{-\tau R})+y_0)^{\frac{2}{3}} d\Omega^2.
\end{equation}
The expression for $K$ is a bit messy, but straightforward to obtain.
The behavior of the function $y(R)$ for these examples depends on the
sign of $\tau$. Recalling that $y_0$ and $y_1$ are both positive, we
find that if $\tau$ is positive, then $y(R)$ has a zero for some
negative value of $R$, and for large $R$ it approaches a constant. If
$\tau$ is negative, then for negative $R$ the function $y$ either hits
zero for some finite value $R_1$ or it approaches zero as
$R\rightarrow - \infty$, while for large positive values of $R$ the
function $y(R)$ is convex increasing and unbounded. Thus in these
cases, either $I=(-\infty, +\infty) $ or $I=(R_1, +\infty).$

Since $R$ is a radial coordinate, which measures distance from the symmetry axis, to obtain a spherically symmetric dynamical horizon from this example, we need to restrict the interval $I$ to $(0,+ \infty)$. When we do this, we find that while the metric is not degenerate as $R\rightarrow 0$, the curvature is unbounded as one approaches the axis. 

Note that since $\frac{dy}{dR}=y_1 e^{-R \tau} >0$ for any constant $\tau$, the (DH3) condition is satisfied for the full range of $I$.

\paragraph{Flat 3-metric:} If we set $r(R) =R$, then $(\Sigma,\gamma)\simeq
(\bR^3,\delta)$.  To see which expressions for $K$ and $\rho$ and
$\xi$ are compatible with a flat metric $\gamma=\delta$, we substitute
 $y(R)=R^3$ into \bref{yeq}, giving
\begin{equation} 
6R + 3\tau  R^2-\tfrac{3}{2} \xi R^3 =0.
\end{equation}
Assuming co-moving matter $\xi=0$ leads to $\tau=-\frac{2}{R}$. From \bref{DH}
we find that $\mu=-\frac{2}{R}$, so $K=-R d\Omega^2$ and the
 matter density is $\rho=\frac{1}{R^2}$. The data is clearly singular at $R=0$, and nowhere else. Without the
co-moving assumption, we find that equation \bref{yeq} implies that
\begin{equation}
3R(2 +\tau R -\tfrac{1}{2}\xi R^2)=0.
\end{equation}
Choosing, say, $\xi = \frac{1}{3 R^2}$, gives $\rho =\frac{2}{3 R^2}$,
$\tau= -\frac{11}{6R}$ and $\mu=-2/R$ as before. One then readily
constructs $K$, noting its singular behavior at the symmetry axis.  

For all cases of this example, since $\frac{dy}{dR}=1 >0$, condition (DH3) for a SSDH is satisfied everywhere.

\subsection{Physical Restrictions} \label{sec5}
In the examples just discussed, we have made restrictions on the
choice of the free data $\xi$ and $\tau$ based on mathematical
convenience. In this section, we consider physically motivated
restrictions, and some of their consequences. 

Our first physical condition on a dynamical horizon $\Sigma$ is that
there is at least one point on the symmetry axis, which we may
normalise to $R=0$,
\begin{equation}
  \label{cos}
r(R=0)=0.
\end{equation}
Second, we recall that the definition of dynamical horizon requires
that the foliation 2-surfaces $S^2_R$ should be contracting in the
inward null direction $n=\ell_-$, but that this condition plays no
direct role in the solution of the dynamical horizon equations
(\ref{gamma}--\ref{yeq}).  Thus we consider explicitly the \emph{inner
  trapped} condition
\begin{equation}
  \label{ITS}
 \theta_- < 0.
\end{equation}
As shown by \cite{DM05,CD95}, the inner trapped condition in spherical
symmetry is \emph{non-evolutionary}, meaning that if it is satisfied on a
Cauchy surface then it holds throughout the globally hyperbolic
development.  We can also show that \bref{ITS} follows from the
\emph{outgoing momentum} condition
\begin{equation}
  \label{xipos}
 \xi\ge 0.
\end{equation}
Finally, we recall that the conceptual picture of a dynamical horizon
has it inside an event horizon, with the area function $r$ having a
finite bound $r\le 2m(\infty)$ where $m(\infty)$ is the final Bondi
mass.  The examples above show that this condition is not satisfied by
all solutions of the dynamical horizon constraint equations
\bref{ham},\bref{mom}.  This motivates the \emph{black hole}
condition, that there is a constant $r^*>0$  with
\begin{equation}
  \label{bhdh}
 r^* := \sup_{(0,R^*)}r,
\end{equation}
where $R^*$ is \emph{maximal}, in the following sense:
\begin{Definition} \label{def2}
Given the data $(\tau,\xi,y_1,R^*) \in C^0(I)\times C^0(I)\times (0,\infty)\times
  (0,\infty]$ specified on the interval $I=(0,R^*)$, let the corresponding solution 
$(y,r,\mu,\tau,\rho,\xi)=:\Sigma(\tau,\xi,y_1,R^*)$ be constructed as prescribed in Proposition \ref{prop1} 
with $R_0=0$, $y_0=0$. If this solution  satisfies $y>0$ and $\theta_-<0$ on $I$, then $(\tau,\xi,y_1,R^*)$ is called a
\emph{DH data set}. Such a DH data set is
  \emph{DH-extendible} if there is a DH data set
  $(\hat{\tau},\hat{\xi},\hat{y}_1,\hat{R})$ such that $\hat{R}>R^*$
  and $\hat{\tau}|_{(0,R^)}=\tau$, $\hat{\xi}|_{(0,R^)}=\xi$ and
  $\hat{y}_1=y_1$. A DH data set $(\tau,\xi,y_1,R^*)$ is
  \emph{maximal} if it is not DH-extendible.  If the inner trapped
  condition is not assumed then we have a \emph{generalised DH data
  set}, and a DH data set is \emph{weakly DH-extendible} if it is
  extendible in the class of generalised DH data.
\end{Definition}

\begin{Proposition}\label{prop4}
  Suppose $\Sigma(\tau,\xi,y_1,R^*) = (y,r,\mu,\tau,\rho,\xi)$ is a
  dynamical horizon constructed by Proposition \ref{prop1} on
  $I=(0,R^*)$, so in particular, $\Sigma$ satisfies the axis condition
  \bref{cos}.
\begin{enumerate}
\item[(i)] Suppose $(\tau,\xi,y_1,R^*)$ is a generalised DH data set, so
  the inner trapped condition \bref{ITS} is not assumed a priori.  If
  the outgoing momentum condition \bref{xipos} is satisfied on a
  sub-interval $(0,\tilde{R})$ then the dynamical horizon is inner
  trapped on the same sub-interval, and
  $\lim_{R\uparrow\tilde{R}}\theta_-<0$.
\item[(ii)] Suppose the black hole \bref{bhdh} and inner trapped
  \bref{ITS} conditions hold on $(0,R^*)$, where $R^*$ is maximal.
  Then $r(R)$ is increasing on $(0,R^*)$, $\lim_{R\uparrow R^*}\ddR r
  =0$, and either $R^*<\infty$ and $\theta_-(R^*)=0$, or $R^*=\infty$.
\end{enumerate}
\end{Proposition}

\Proof (i)\ \  It follows from $\theta_+=0$ and $n=\ell-2e_R$ that
$\theta_- = -4 \ddR \log r = -\tfrac{4}{3}\ddR\log y$.  Proposition
\ref{prop1} shows that if $y(0)=0$ then for small positive $R$ we have
$\theta_- \simeq -\tfrac{4}{3} y'(0)/R <0$.  The master equation
\bref{yeq} and $\theta_+=0$ show that on $\Sigma$ we have
\begin{equation}
  \label{eq:2}
  \ddR \theta_- = \tfrac{3}{4} \theta_-^2 -\tau\theta_- -  2\xi
 \ \ \le \ \ \tfrac{3}{4} \theta_-^2 -\tau\theta_-\;,
\end{equation}
since $\xi\ge0$ by assumption. Introducing $h=-1/\theta_-$ and
integrating gives
\[
h(R) \le e^{T(R)}\left( h(\epsilon) + \frac{3}{4} \int_\epsilon^R
  e^{-T(s)}\,ds\right),
\]
where $T(R)= \int_\epsilon^R \tau(s)\,ds$ and $\epsilon>0$ is chosen
small enough that $h(\epsilon)\le 1$.  Now $\tau\in C^0(\bR)$ so
$h(R)$ is bounded above uniformly for finite $R$ and thus
$\theta_-=-1/h$ is negative as required; even more, $\theta_-$ is
locally uniformly negative, since $\int_R^{R+1}\tau(s)\,ds$ is
bounded, for all $R$.
\\
(ii)\ \ The inner trapped condition $\theta_-<0$ and $\theta_-=-4 \ddR
\log r$ shows that $\ddR r>0$.  The master equation \bref{yeq} and
$\tau,\xi\in C^0(\bR)$ show $\ddR y=3r^2 \ddR r$ is bounded locally
uniformly on $\bR$, so $\lim_{R\uparrow R^*}\ddR r < \infty$ while $r$
is bounded away from $0$.  Choosing any $C^0$ extension
$(\tilde{\tau},\tilde{\xi})$ of $(\tau,\xi)$ gives an $\tilde{R}>R^*$
and a generalised DH data set
$(\tilde{\tau},\tilde{\xi},y_1,\tilde{R})$ extending $\Sigma$, by
continuity of $y$ at $R^*$.  Continuity of $\ddR y$ implies that
$\Sigma$ is DH-extendible, contradicting maximality of $R^*$, unless
$\lim_{R\uparrow R^*}\ddR r =0$.  Finally, if $R^*<\infty$ then
continuity of $\ddR y$ shows that $\theta_-(R^*)=0$.  \QED

\Remark Proposition \ref{prop4} provides a natural picture of a
dynamical horizon inside an event horizon, starting at the central
axis and extending radial-outward and geodesically complete, with
future endpoint at $i^+$.  However, this picture relies on the
assumptions we have made, in particular the inner trapped condition
\bref{ITS} and the continuity of the free fields $\tau,\xi$.  Examples
of marginally trapped tubes (cf. \cite{AG05,BBGV05}) show that the
condition $\theta_+=0$ defines a hypersurface which may become null
and timelike.  In these examples, the dynamical horizon metric is
inextendible as a spacelike hypersurface but not complete.  The next
results examine the solutions near the singular points $0$, $R^*$.





\begin{Lemma}\label{lemma2}
  Suppose $\Sigma=\Sigma(\tau,\xi,y_1,R^*)$ is a spherically symmetric
  dynamical horizon with $R^*<\infty$.  Let $(E_R,E_\perp)$ be a frame
  along $\Sigma$ which is parallel transported by the spacetime
  connection.  Then there is a finite boost (with parameter $a$) to
  the $\Sigma$-adapted frame $(e_R, e_\perp)$ at $R=R^*$,
\begin{eqnarray*}
e_R & =& \half (a-a^{-1}) E_\perp +\half(a+a^{-1}) E_R,
\\
e_\perp &=& \half (a+a^{-1}) E_\perp +\half (a-a^{-1}) E_R.
\end{eqnarray*}
Furthermore, $\Sigma$ has spacelike radial unit tangent
vector $e_R$ at $R=R^*$ and if $\lim_{R\uparrow R^*} \theta_-<0$ then
$\Sigma$ is DH-extendible.
\end{Lemma}  

\Remark The spacetime parallel transport provides a reference frame,
uniformly equivalent to any other construction of spacetime frame, and
therefore suitable for verifying whether or not $e_R$ goes null at
$R^*$.

\Proof The extrinsic curvature $K_{ab}$ is defined by $K(X,Y)=
g(X,\nabla_Y e_\perp)$, so $\nabla_{e_R} e_R = K(e_R,e_R)e_\perp$ and
$\nabla_{e_R} e_\perp = K(e_R,e_R)e_R$.  The parallel frame
$(E_R,E_\perp)$ satisfies $\nabla_{e_R}E_R = \nabla_{e_R}E_\perp =0$.
The corresponding null frames $\ell_\pm = e_\perp\pm e_R$ and $L_\pm =
E_\perp\pm E_R$ are related by a boost with parameter $a=a(R)$,
ie.~$\ell_\pm = a^{\pm1}L_\pm$, and we find $\ddR \log a =
K(e_R,e_R)=\tau-\mu =\tau+2\ddR\log r$.  Integrating gives
\begin{equation}\label{boost}
\frac{a(R)}{r(R)^2} = \frac{a(R_1)}{r(R_1)^2} \exp\Bigl(\int_{R_1}^R
\tau(s)\,ds\Bigr),
\end{equation}
so the boost $a(R)$ is bounded while $r>0$ and $\int \tau <\infty$.
Now $\theta_-(R^*)<0$ and $\theta_- = -4\ddR\log r$ show that $\ddR
r(R^*)>0$, so the dynamical horizon solution extends as before.
\QED

A similar argument can be used to show that the Einstein tensor is
unbounded at the axis.

\begin{Corollary}\label{cor1}
  Suppose $\Sigma$ is a spherically symmetric dynamical horizon in a
  spacetime $M$, which satisfies the axis condition \bref{cos} and
  $\int_0^1 \tau(s)\,ds <\infty$.  Then the central axis point $R=0$,
  $r=r(R=0)=0$ of $\Sigma$ lies on the singular set of the spacetime
  $M$.
\end{Corollary}
\Proof
Equation \bref{7} shows that the component $G(e_\perp,\ell)$ is
unbounded, and we must show this holds in all frames near $R=0$.  It suffices
 to consider frames $L,N$ which are spacetime parallel
along $\Sigma$.  As in Lemma \ref{lemma2} the tangent frames $n,\ell$
and parallel frame $N,L$ are related by a boost with parameter
$a=a(R)$; i.e., ~$\ell=aL$ and $n=a^{-1}N$, where $a=r^2O(1)$ as $R\to0$ by
\bref{boost}. It follows from $r^{-2}=\half(G(\ell,n)+G(\ell,\ell))=
\half(G(L,N)+a^2G(L,L))$ that $G(L,L)=O(r^{-6})$ is the required
unbounded coefficient.  Note also that the boost $n=O(r^{-2})N$ so
$\Sigma$ is tangent to the past null cone at the central axis point;
see (\cite{BBGV05},Figure 3.). 
\QED


%
%

\section{Spacetime Picture}\label{sec4}

Let $(M,g, T)$ be a spherically symmetric spacetime which satisfies
the Einstein equations $G_{\mu \nu}=8\pi \newton\,T_{\mu\nu}$ with
stress-energy $T_{\mu\nu}$, and consider the question of finding a
dynamical horizon in $M$. If one were to exist in $(M,g, T)$, then the
initial data on this dynamical horizon would necessarily satisfy the conditions
discussed in Section \ref{sec3}, including \bref{7}. Since \bref{7}
depends on the adapted frame $(e_\perp ,e_R)$, it is not easy to check
directly from the spacetime fields. One can however state an
equivalent slice-independent condition:
\begin{Lemma} \label{lem1}
Let $W\subseteq M$ be a spacetime region with stress-energy tensor $T$
satisfying the strict null energy condition (SNEC)
\begin{equation}
\label{SNEC}
T(L,L)> 0 \textrm{ for all radial null vectors } L \textrm{ in } W.
\end{equation}
  Then the following are equivalent:
\begin{itemize}
\item[(i)] There is a future null frame $(\ell,n)$ with $g(\ell,n)={}-2$
satisfying the condition
\begin{equation}
\label{matter2}
\half r^2 G(n,\ell) = 4\pi \newton r^2 T(n,\ell) < 1.
\end{equation}
\item[(ii)] There is a spacetime frame $(E_\perp,E_R)$ satisfying
\bref{7}.
\end{itemize}
\end{Lemma}

\Remark  The Lorentz-invariant condition \bref{matter2} arises
frequently: for example, in
\cite{BBGV05} the sign of the \emph{C-function} (\cite{BBGV05}
equation (2.3))
\begin{equation} \label{Cfn}
C = \frac{\half r^2 G(\ell,\ell)}{1-\half r^2 G(\ell,n)}
\end{equation}
(or more accurately, the sign of $C^{-1}$) determines the causal
character (spacelike/null/timelike) of the marginally trapped tube
(MTT).  Another example is the evolution equation \cite{DM05}
\begin{equation}\label{r-hyp}
\frac{\partial^2 (r^2)}{\partial u\partial v} = -\Omega^2(1-\half
G(\ell,n)r^2)
\end{equation}
for the areal function $r^2(u,v)$ in double null coordinates
$ds^2=-\Omega^2(u,v)\,du\,dv+r^2(d\theta^2\sin^2(\theta)\,d\phi^2)$.

\Proof To show that $(i)\Rightarrow(ii)$, we check the effect of a
Lorentz boost on $T(e_\perp,\ell)$.  The boosted frame $L=a\ell$,
$N=a^{-1}n$, $E_\perp=\half(L+N)=\half(a\ell+a^{-1}n)$ satisfies
\[ 
2T(E_\perp,L) = T(N,L) + a^2 T(L,L)\,;
\]
hence it follows from the SNEC \bref{SNEC} and the Lorentz invariance
of \bref{matter2} that there is a unique $a>0$ such that
$T(E_\perp,L)=\half (4\pi \newton  r^2)^{-1}$, which is equivalent
to the condition \bref{7}.  The converse $(ii)\Rightarrow(i)$ follows
from the above boost relation, together with the definitions of $\rho$
and $\xi$ in terms of $T$.  \QED

It follows from this Lemma that, if we are given a spacetime which
satisfies the SNEC, then inequality \bref{matter2} is a necessary
condition for it to contain a dynamical horizon. We now look to find
sufficient conditions for existence.
Suppose that in
our given spacetime $(M,g,T)$ there is a non-empty region $W\subseteq
M$ in which the SNEC condition and \bref{matter2} both hold.  From the
lemma, it follows that through every point in $W$ there is a spacelike
vector $e_R$ such that the adapted frame $(e_\perp,e_R)$ satisfies
\bref{7}.  The integral curves of the vector field $e_R$ give a
foliation of $W$ by spacelike hypersurfaces which satisfy \bref{7}. It
follows from the results of Section \ref{sec3} that if the spacetime
contains a dynamical horizon, it must coincide with a leaf of this
foliation.

To see that condition \bref{7}, whilst necessary for a given leaf to
be a dynamical horizon, is not sufficient\footnote{We have shown in
  Section \ref{sec3} that while \bref{7} holds, for any choice of
  a pair of free functions, we can construct initial data so that the
  hypersurface with that data is a dynamical horizon for the spacetime
  development of that data. It does \emph{not}, however, follow that
  any initial data set satisfying \bref{7} is a dynamical horizon: we
  show here that the additional condition $\theta_+(p)=0$ for at least
  one point is necessary.}, we use the constraint
equations with \bref{7} and the expression $\theta_+ =
\mu+\frac{2}{r}\ddR r$ for the future outer expansion $\theta_+$ to
derive a differential equation for $\theta_+$.  Specifically, writing
\bref{mom} in terms of $\theta_+$ and eliminating $\tfrac{d^2 r}{dR^2}$
with \bref{ham}, we obtain
\begin{equation} \label{theq}
\frac{d}{dR}\theta_+ + (\tfrac{3}{4}\theta_+ -\tau)\theta_+
= \frac{1}{r^2} - \rho-\xi \,.
\end{equation}
Clearly if we choose a spacelike hypersurface with adapted frame
$\{e_\perp, e_R\}$ relative to which $\rho +\xi=\frac{1}{r^2}$ (which
Lemma 2 guarantees that we always can do) then the right hand side of
\bref{theq} vanishes, and we see that \bref{theq} admits the solution
$\theta_+=0$. It then follows from ODE uniqueness that the sign of
$\theta_+$ is fixed on each leaf.  Hence if $p\in W$ and
$\theta_+(p)=0$ then the leaf through $p$ is a dynamical horizon.
Finally, it follows from the Raychaudhuri equation
\[
D_\ell \theta_+ = - \half\theta_+^2 - 8\pi \newton T(\ell,\ell)
\]
and from the strict null energy condition
that $D_\ell\theta_+ < 0$. Presuming the spacetime to be smooth, we
find that there are four possibilities:
\begin{enumerate}
\item[(a)] The region $W$ has $\theta_+>0$
everywhere, so that it is entirely untrapped, and there is no
dynamical horizon. 
\item[(b)] The region $W$ has $\theta_+<0$ everywhere, so
that it is entirely trapped, and there is no dynamical horizon.
\item[(c)]  The region $W$
contains a single leaf with $\theta_+=0$ , but $\theta_-\geq 0$ everywhere on this leaf. 
So there is no dynamical horizon.
\item[(d)] The region $W$ contains a single leaf with $\theta_+=0$, and on some (possibly proper)
subset of this leaf, $\theta_- < 0$. This subset is a dynamical horizon. 
\end{enumerate}
In summary, we have shown the following:
\begin{Proposition} \label{prop3}
Let $(M,g,T)$ be a smooth spacetime which satisfies the strict null
energy condition \bref{SNEC}.
\begin{enumerate}
\item If condition \bref{matter2} is satisfied nowhere in $M$, then
  the spacetime contains no dynamical horizons.
 \item If there exists a non-empty connected region $W\subseteq M$ in
  which condition \bref{matter2} holds, and if the outward null
  expansion function $\theta_+$ has constant sign (with no zeroes)
  then $W$ contains no dynamical horizon and is everywhere trapping or
  non-trapping.
\item If there exists a non-empty connected region $W\subseteq M$ in
  which condition \bref{matter2} holds, and if the expansion
  $\theta_+$ takes on both positive and negative values in $W$, then
  $W$ contains a unique spacelike hypersurface with $\theta_+=0$.  The
  data on this hypersurface satisfy \bref{7}, and it is (entirely or
  partly) a dynamical horizon if and only if $\theta_-<0$ on all of it
  or part of it. In addition, this hypersurface divides $W$ into a
  trapping region to the future and a non-trapping region to the past.
\end{enumerate}
\end{Proposition}
\Remark It is worth noting that, unlike the situation for dynamical
horizons in non-spherical spacetimes, the spherically symmetric
construction given here does not depend on any choice of (spherically
symmetric) time slicing.  This leads to a much stronger uniqueness
statement (Proposition \ref{prop3}) than is available in the general
case.

\section{Conclusion}

While the set of spacetimes which are spherically symmetric is very
special, it is clear from physical considerations that, so long as
matter is present, dynamical horizons can form in them. We determine
here the necessary conditions for spherically symmetric spacetime to
contain a dynamical horizon, and sufficient conditions as well. We
also demonstrate their uniqueness, and show how to systematically
construct initial data sets which serve as dynamical horizons for
their spacetime developments (with two free functions parametrizing
the collection of all such data sets).

The situation regarding dynamical horizons in non-spherically
symmetric spacetimes, is of course considerably more complicated.
Still, a full understanding of the spherically symmetric case provides
a good first step toward uncovering the properties of dynamical
horizons more generally.
 
\section{Acknowledgments} 
We thank Abhay Ashtekar, Greg Galloway and Dan Pollack for useful
discussions, and the Isaac Newton Institute and the Australian
National University for hospitality at various stages. The work of JI
is partially supported by NSF grant PHY-0354659 at Oregon, and RB
acknowledges support from the Clay Institute and the Australian
Research Council.

\end{document}